\documentclass[apl,showpacs,twocolumn]{revtex4}
\usepackage{amsfonts}
\usepackage{amsmath}
\usepackage{amssymb}
\usepackage{graphicx}
\usepackage{subfigure}

\begin{document}

\title{\emph{Ab initio} investigation on oxygen defect clusters in UO$_{2+x}$}
\author{Hua Y. Geng}
\affiliation{Department of Systems Innovation, The University of Tokyo, Hongo 7-3-1, Tokyo 113-8656,
Japan}
\author{Ying Chen}
\affiliation{Department of Systems Innovation, The University of Tokyo, Hongo 7-3-1, Tokyo 113-8656,
Japan}
\author{Yasunori Kaneta}
\affiliation{Department of Systems Innovation, The University of Tokyo, Hongo 7-3-1, Tokyo 113-8656,
Japan}
\author{Motoyasu Kinoshita}
\affiliation{Nuclear Technology Research Laboratory, Central Research Institute of Electric Power Industry, Tokyo 201-8511, Japan}
\affiliation{Japan Atomic Energy Agency, Ibaraki 319-1195, Japan}

\begin{abstract}
By first-principles
LSDA+U calculations, we revealed that the current
physical picture of defective uranium dioxide suggested solely by neutron diffraction analysis
is unsatisfactory.
An understanding based on quantum theory has been established as a thermodynamical competition among point defects and cuboctahedral cluster,
which naturally interprets the puzzled origin of the asymmetric O$^{'}$ and O$^{''}$ interstitials.
It also gives a clear and consistent agreement with most available experimental data.
Unfortunately, the observed high occupation of O$^{''}$ site cannot
be accounted for in this picture and is still a challenge for theoretical
simulations.
\end{abstract}

\volumeyear{year}
\volumenumber{number}
\issuenumber{number}
\eid{identifier}
\maketitle


Fluorite structure commonly appears
in rare earth and actinide oxides. One of them UO$_{2}$ is the nowadays widely adopted nuclear
fuel and most of whose applications are
highly related to atomic defect behaviors.
The current physical picture of defective UO$_{2}$ is based purely on neutron diffraction measurements.
It suggested that no oxygen can occupy the octahedral site (O$_{i}$) of the cation FCC lattice
and all oxygen interstitials should displace from this site along the $\langle110\rangle$ and $\langle111\rangle$
directions about 1\,\AA (sites O$^{'}$ and O$^{''}$), respectively, to form the so-called \emph{Willis type}
\emph{clusters} by associating with the nearby oxygen vacancies (O$_{v}$).\cite{willis64a,willis64b,willis78,murray90}
Note the clustering interpretation in this picture is suggestive and might be ambiguous since explicit atomic positions
are unavailable in those experiments.
But this suggestion invalidated
the simple point defect model and diffusion mechanism that widely employed to describe
fuel behaviors.\cite{matzke87,lidiard66,geng08} Also itself is inconsistent in that it requires a quite different
cuboctahedral (COT) cluster to account for the closely related U$_{4}$O$_{9}$/U$_{3}$O$_{7}$ phase.\cite{bevan86,cooper04,garrido03,nowicki00,
willis64b,allen82}
This situation provoked confusion because it is difficult
to investigate the Willis clusters thoroughly and meanwhile
people are unsure how large the clustering effects should be.\cite{catlow77,govers07,geng08,crocombette01,freyss05}

On the other hand, as a powerful quantum mechanics technology, density
functional theory provides an accurate \emph{ab initio} method to
understand material behaviors.\cite{kresse96}
For example the bulk stoichiometric UO$_{2}$ has been well described by
local density approximation with Hubbard correction (LSDA+U\cite{anisimov91}) functional.\cite{geng07}
Application of this method to non-stoichiometric UO$_{2+x}$
also showed amazing agreement with available experiments,
namely, in ({\romannumeral 1}) negative volume change induced by oxygen interstitials,
({\romannumeral 2}) predominance of oxygen defects when
$x>0$, ({\romannumeral 3}) exclusive COT clusters at high $x$
region (U$_{4}$O$_{9}$/U$_{3}$O$_{7}$).\cite{willis64a,willis64b,willis78,murray90,
matzke87,lidiard66,geng08,bevan86,cooper04,garrido03,geng08b}
In this letter we will investigate the full temperature-composition region
to study the oxygen defect clustering behaviors in UO$_{2+x}$ with this method.

All defects were modeled in a $2\times2\times2$ fluorite cubic cell with otherwise 96 atoms.
Larger supercells also tried to check the size effect. Structure optimization
has been performed with P1 symmetry and until
all forces and stress less than 0.01\,eV/\AA.
The employed computational technique and parameters, such as the
kinetic energy cutoff and sampling k-points, are
the same as in Ref.[\onlinecite{geng08}].

\begin{figure}[h]
\centering
\includegraphics[width=3.5 in]{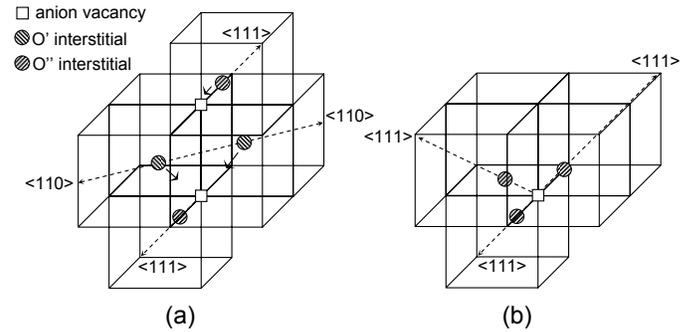}
  \caption{Relaxation process of the Willis 2:2:2 cluster (a) to a $V$-3O$^{''}$ split interstitial (b):
  each cube indicates one oxygen cage and the solid arrows point to atomic relaxation directions;
  in (a) the topside interstitial will annihilate the nearest vacancy eventually and the two O$^{'}$
  interstitials move downwards to the nearby O$^{''}$ sites.}
  \label{fig:struct}
\end{figure}

\begin{figure}[h]
  \includegraphics*[width=2.5 in]{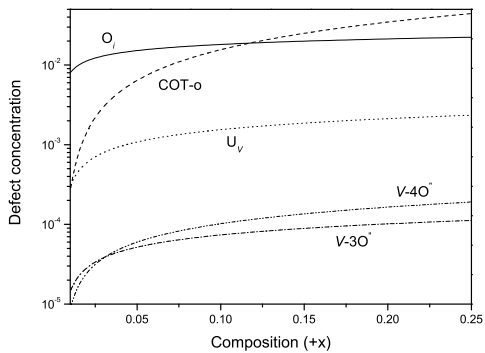}
  \caption{Defect concentrations of point oxygen interstitial, uranium vacancy,
  COT-o, $V$-3O$^{''}$ and $V$-4O$^{''}$ clusters at 1500\,K, respectively. All other components are negligible.
}
  \label{fig:deft-conc1}
\end{figure}

Willis-type cluster has a geometry of that O$^{'}$ interstitials
pushing the nearest lattice oxygen displaced
along the $\langle111\rangle$ direction
and thus created a connected O$^{'}$:O$_{v}$:O$^{''}$ defect cluster.\cite{willis64b,willis78}
A typical 2:2:2 cluster is shown in figure \ref{fig:struct}(a).
Other already suggested
clusters include 1:2:2, 4:3:2, 2:3:2 and so on.\cite{catlow77}
We have checked the structural stability
of several simple clustering models
in Ref.\cite{geng08}, but
failed to give a definite conclusion for Willis clusters due to the limitation of the
method employed there.
Here we will investigate the stability of 1:2:2, 2:2:2 and 3:3:2 (with 2 inequivalent configurations, respectively),
4:3:2, and 2:3:2 (with 3 inequivalent configurations) clusters directly.
The behaviors of large Willis clusters can be inferred from these small ones.
Calculations showed that for each kind of cluster, all inequivalent configurations
always relaxed to the same final structure. In details, all 1:2:2 clusters relaxed
to a point oxygen interstitial occupying the octahedral site (O$_{i}$),
all 2:2:2 and 3:3:2 clusters relaxed to a $V$-3O$^{''}$ cluster, the 4:3:2
relaxed to a $V$-4O$^{''}$ cluster, and all 2:3:2 decayed to point O$_{i}$.
There is no energy trap presented along the relaxation paths, showing
that Willis-type cluster is not even metastable.

Figure \ref{fig:struct}(a) shows the relaxation process of a
2:2:2 cluster, in which the arrows indicate the atomic movement directions.
Figure \ref{fig:struct}(b) gives the final $V$-3O$^{''}$ cluster.
The failure of Willis-type clusters is owing to that O$^{'}$ interstitial in fact cannot
push the nearest lattice oxygen out along the $\langle111\rangle$ direction.
Instead itself will relax to an O$^{''}$ site to form a split-interstitial ($V$-3O$^{''}$/$V$-4O$^{''}$)\cite{geng08}
or to annihilate nearby vacancy.
This is different from that in COT cluster where the integral support
stabilizes the O$^{'}$ interstitials.\cite{geng08b}
In the case of 2:2:2, the topside O$^{''}$ will annihilate
the nearest O$_{v}$, with a little distortion along the $\langle001\rangle$ direction;
the two O$^{'}$s will move downwards to the nearest O$^{''}$ sites and form
a $V$-3O$^{''}$ cluster with the low O$^{''}$ interstitial.
This rule applies to other Willis clusters too. For example, In 1:2:2 and
2:3:2 cases, O$^{'}$ interstitials were not capable of supporting the structure
and all interstitials annihilated nearby vacancies with the last one
relaxed to an O$_{i}$ site; the four O$^{'}$s in 4:3:2
also couldnot prevent the O$^{''}$s from annihilating
and themselves then relaxed to O$^{''}$ positions surrounding the middle
vacancy and formed a $V$-4O$^{''}$ cluster.

This makes the concept of Willis cluster generally failed.
For clusters bigger than 4:3:2, there is no long-ranged interaction that can stabilize
the structure. We have tried several primary calculations on big clusters
but failed to find any evidence that they were favored.

\begin{table}[h]
\caption{\label{tab:Struct-E} First principles results for oxygen defect clusters
in UO$_{2+x}$:
$\Delta V$ the
defect induced
volume change per fluorite cubic cell;
E$_{f}$, E$_{ef}$ and E$_{pf}$ are the
overall defect formation energy, formation energy per excess oxygen and per oxygen Frenkel pair, respectively.
}
\begin{ruledtabular}
\begin{tabular}{l c c c c c c c c c c} 
   & $x$ & $\Delta V$(\AA$^{3}$) & E$_{f}$(eV) & E$_{ef}$(eV) & E$_{pf}$(eV) \\
\hline
  $V$-3O$^{''}$  & $\frac{1}{16}$   & $-0.21$      &$-3.88$&$-1.94 $&$3.72$  \\
  $V$-4O$^{''}$  &$\frac{3}{32}$  & $-0.42$        &$-6.74$&$-2.25 $&$3.96$\\
\end{tabular}
\end{ruledtabular}
\end{table}

Since all Willis-type clusters will relax to O$_{i}$, $V$-3O$^{''}$,
$V$-4O$^{''}$ and their combinations, it is reasonable to assume that
$V$-3O$^{''}$ and $V$-4O$^{''}$ clusters might
take the role of Willis clusters to account for the measured occupation number of O$^{'}$
and O$^{''}$ sites.\cite{willis64b,willis78,murray90}
To verify this assumption,
we calculated these two clusters in a $2\times2\times2$
supercell. The defect induced volume changes and energetic information are listed
in table \ref{tab:Struct-E}.
Like O$_{i}$ and COT clusters, $V$-3O$^{''}$ and $V$-4O$^{''}$ clusters
all result in a negative volume change, but is much smaller than that of COT-o (COT
with one O$^{''}$ occupies the center).\cite{geng08,geng08b}
As suggested by counting the nearest vacancy-interstitial pairs,\cite{geng08} $V$-4O$^{''}$ has
lower overall and per excess oxygen formation energies. But due to the
number of Frenkel pair available when compensating with point
O$_{v}$, the formation energy per Frenkel pair of $V$-4O$^{''}$
is a little higher than that of $V$-3O$^{''}$.

The defect concentrations as a function of temperature and composition $x$
were calculated with the independent clusters approximation.\cite{geng08,geng08b}
This method is exact at low temperatures low defect concentrations where atomic vibrational
effects and correlation among defect clusters are negligible.
Under the constraint of
$x=2([V_{U}]-[I_{U}])+[I_{O}]+\sum_{i}n_{i}\rho_{i}+2\sum_{j}n_{j}\rho_{j}-2[V_{O}]$
where $i$ runs over COT-v and COT-o clusters\cite{geng08b} and $j$ over $V$-3O$^{''}$
and $V$-4O$^{''}$ (with $n$ is the excess number of oxygen in each cluster, and the coefficient
2 before the second summation arises from the fact that these clusters are
defined on the oxygen sublattice), we then got the defect concentrations by Eq.(1)
of Ref.\cite{geng08b} in the closed regime where no particle-exchange with the exterior occurs.



The results at 1500\,K are shown in figure \ref{fig:deft-conc1},
where all unshown defect concentrations are small enough to be ignored.
Cluster $V$-3O$^{''}$ and $V$-4O$^{''}$
compete each other throughout the whole $x$ region,
while their concentrations are always 1 order smaller
than uranium vacancy and 2 orders smaller than that of O$_{i}$ and COT-o.
At lower
temperatures their concentrations reduce further, and can be neglected.

In this way a physical picture about oxygen defects appears, \emph{i.e.},
the nontrivial oxygen interstitials in hyperstoichiometric regime
are only O$_{i}$ and COT-o.
All O$^{''}$ and O$^{'}$ should come from
COT-o clusters.
This picture describes very well the high $x$ region where U$_{4}$O$_{9}$/U$_{3}$O$_{7}$ dominated.\cite{geng08b}
It also explains the origin of asymmetric O$^{'}$ and O$^{''}$
interstitials and the non-negligible O$_{v}$ at T$\sim300$\,K and $x\sim0.1$ qualitatively.
However, it left a quantitative discrepancy:
the experimental occupation number of O$^{'}$ and O$^{''}$
is 0.08:0.16\cite{willis64b} (which had been corrected to 0.13:0.12 late\cite{willis78}), 0.14:0.12, and 0.33:0.10\cite{murray90} when $x$ at around 0.11$-$0.13.
It evidently asks for a low O$^{'}$:O$^{''}$ ratio, but COT-o
cluster can provide only a ratio of 12:1.\cite{geng08b}

This disagreement is unlikely due to the failure of the current model.
Three independent measurements on U$_{4}$O$_{9}$ and U$_{3}$O$_{7}$ has definitely
proved that the clusters in these phases are COT exclusively, \emph{i.e.},
the O$^{'}$:O$^{''}$ ratio
should approach to 12:1 at high $x$ region.
Meanwhile, the monotonicity of defect concentrations
\emph{vs} $x$ is insensitive to the exact formation energies.
It seems impossible that $V$-3O$^{''}$/$V$-4O$^{''}$ cluster (contributor to O$^{''}$)
would have a considerable concentration at $x\simeq0.11$ but disappear at a higher one.
However, the possible decomposition/reassembly of COT clusters might have
some impact on this issue, and we look forward to some understanding from
this aspect.

On the experiment side, ambiguity existed in its post data analysis.
Recall the correction of the O$^{'}$:O$^{''}$ ratio
from 0.08:0.16 to 0.13:0.12\cite{willis64b,willis78} and the similar vagueness
in U$_{4}$O$_{9}$/U$_{3}$O$_{7}$ about the position of the
center oxygen in COT clusters,\cite{bevan86,cooper04,garrido03,geng08b}
we know it is very difficult
to extract the explicit atomic geometry from powder neutron diffraction measurements, especially
for UO$_{2+x}$ that is not an ordered phase and some peaks were broadened drastically.
Another cause that might affect
the O$^{''}$ occupation is oxygen partial pressure.
If experiment were performed
in an open atmosphere and the powder were fine enough, then the
above model need to be revised to take oxygen-exchange
with the exterior into account.
Then the defect concentration becomes proportion to\cite{crocombette01}
$\rho_{i}\sim \exp\left(-\frac{E_{f}^{i}}{\kappa_{B} T}\right)\left(P_{\mathrm{O}_{2}}f(T)\right)^{n_{i}/2}$
where oxygen partial pressure $P_{\mathrm{O}_{2}}$ could modify the concentration
of $V$-3O$^{''}$/$V$-4O$^{''}$ cluster greatly.
Anyway, it requires more endeavor from theorist and experimentalist
to remove this discrepancy.



\begin{figure}[h]
  \includegraphics*[width=2.5 in]{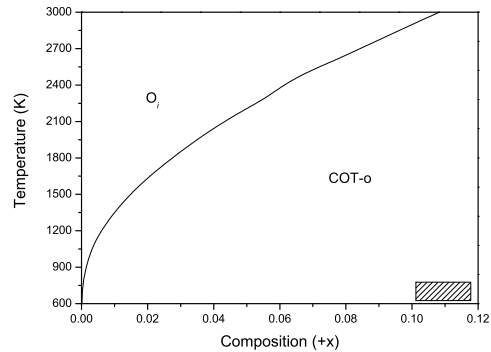}
  \caption{Pseudo phase diagram for oxygen interstitials in UO$_{2+x}$,
  where the hatched area marks the region in which the neutron diffraction measurements
  performed.}
  \label{fig:deft-conc2}
\end{figure}

In summary we clarified the defect clustering structure in UO$_{2+x}$ by
first principles calculations. The Willis clusters have been proved generally failed.
A picture of oxygen defective UO$_{2}$ was established based on thermodynamical
competition between O$_{i}$ and COT-o cluster, which explains most available
experiments very well.
The last confusion is about the role of O$_{i}$ in UO$_{2+x}$.
Willis once argued that oxygen should not occupy this site based on his observations.\cite{willis64a,willis64b}
Above discussion indicated this argument is inappropriate. In order to explain the
Willis' observation that O$_{i}$ is negligible we should take the effect of temperature
and composition into account. Using the calculated defect concentrations,
we plotted a pseudo phase diagram for oxygen defects in UO$_{2+x}$, as shown in figure \ref{fig:deft-conc2}.
Note each COT-o cluster contributes 13 single oxygen interstitials and the figure
was plotted according to which kind of interstitial is majority (from O$_{i}$ or from
COT-o). The figure clearly illustrated the predominant region of O$_{i}$ is at low $x$ and high
temperatures. Mind all experiments available so far were conducted in the region marked
by the hatched area,\cite{willis64b,willis78,murray90} which is far from the O$_{i}$ territory. It is then understandable
why experiments failed to detect O$_{i}$.

Support from the Budget for
Nuclear Research of the Ministry of Education, Culture, Sports,
Science and Technology of Japan, based on the screening and counseling by the
Atomic Energy Commission is acknowledged.


\end{document}